\documentclass[12pt,preprint]{aastex}

\def\la{\mathrel{\hbox{\rlap{\hbox{\lower4pt\hbox{$\sim$}}}\hbox{$<$}}}}
\def\sun{\hbox{$\odot$}}

\begin{document}
\title{Deformed Distance Duality Relations and Supernovae Dimming}
\author{J. A. S. Lima\altaffilmark{1,a},  J. V. Cunha\altaffilmark{2,b}, V. T. Zanchin\altaffilmark{3,b}
\affil{$^{a}$Departamento de Astronomia, Universidade de S\~ao
Paulo, USP,
\\ 05508-900 S\~ao Paulo, SP, Brazil\\
$^{b}$Centro de Ci\^encias Naturais e Humanas, Universidade Federal
do ABC\\ 
Rua Santa Ad\'elia 166, 09210 - 170, Santo Andr\'e, SP,
Brazil}}
\altaffiltext{1}{limajas@astro.iag.usp.br}
\altaffiltext{2}{cunhajv@astro.iag.usp.br}
\altaffiltext{3}{zanchin@ufabc.edu.br}







\begin{abstract}
The basic cosmological distances are linked by the Etherington cosmic distance duality relation, $\eta (z) = D_{L}(z)(1+z)^{-2}/D_{A}(z) \equiv 1$, where $D_{L}$ and $D_{A}$ are, respectively, the luminosity and angular diameter distances. In order to test its validity, some authors have proposed phenomenological expressions for $\eta(z)$ thereby deforming the original Etherington's relation and comparing the resulting expressions with the available and future cosmological data. The relevance of such studies is unquestionable since any violation of the cosmic distance duality relation could be the signal of new physics or non-negligible astrophysical effects in the usually assumed perfectly transparent Universe.  

In this letter, we show that under certain conditions such expressions can  be derived from a more fundamental approach  with the parameters appearing in the $\eta(z)$ expression defining the  cosmic absorption parameter as recently discussed by Chen and Kantowski.  Explicit examples involving four different parametrizations  of the deformation function are given. Based on such an approach, it is also found that the latest Supernova data can also be explained in the framework of a pure cold dark matter model (Einstein-de Sitter).  Two different scenarios with cosmic absorption are discussed. Only if the cosmic opacity is fully negligible, the description of an accelerating Universe powered by dark energy  or some alternative 
gravity theory must be invoked. 
\end{abstract}





\section{Introduction}

The \textit{cosmic distance duality relation (CDDR)} is a mathematical identity relating the \textit{luminosity distance} $D_{\scriptstyle
L}$ with the \textit{angular diameter distance} $D_{\scriptstyle A}$
by the expression:
\begin{equation}
  \frac{D_{\scriptstyle L}}{D_{\scriptstyle A}}{(1+z)}^{-2}=1.
  \label{rec1}
\end{equation}
The validity of this constraint uniting the two basic distances in
cosmology  depends neither on the Einstein field equations nor on the
nature of the matter-energy content. It only requires the phase space conservation
of photons and that sources and observers are
connected by null geodesics in a Riemannian spacetime.  Therefore,
it remains valid for spatially homogeneous and isotropic (anisotropic)
cosmologies, as well as for inhomogeneous cosmological models
(Etherington 1933, Basset and Kunz 2004).

The above relation is usually taken for granted when relativistic models of
the Universe are confronted to the existing cosmological
observations. Despite that, the distance-duality relation is in
principle testable by means of astronomical observations. The basic
idea is to find cosmological sources whose intrinsic luminosities
are known (standard candles) as well as their intrinsic sizes
(standard rulers). After determining  both $D_{\scriptstyle L}$ and
$D_{\scriptstyle A}$ at the same redshift it should be possible to
test directly the Etherington result. Naturally, 
by cosmological sources with known $D_{L}$ and $D_{A}$ we are not referring necessarily to 
the same class of objects. Under certain conditions, as recently discussed by Holanda et al. (2010),  
one may consider two different classes of objects as, for instance,  supernovae and 
galaxy clusters for which $D_L$ and $D_A$ are separately determined. Note also that ideally both
quantities must be measured in such a way that  any
relationship coming from specific cosmological models are not used, that is, they must be
determined by means of intrinsic astrophysically measured quantities. In practice, the validity of the CDDR has been  
tested by assuming  a phenomenological deformed expression of the form (Holanda et al. 2010, Meng et al. 2011, Nair et al. 2011, 
Khedekar \& Chakaborti 2011, Gon\c{c}alvez et al. 2011):
\begin{equation}
\frac{D_{\scriptstyle L}}{D_{\scriptstyle A}}{(1+z)}^{-2}= \eta (z),
\label{receta}
\end{equation}
where $\eta(z)$  is the  deformation function which  quantifies a possible epoch-dependent departure
from the standard photon conserving scenario ($\eta=1$).

As it appears,  a deformed CDDR can also be  adopted to test the possibility of a new physics. In this line, Basset \& Kunz (2004) used  supernovae type Ia (SNe Ia) data as
measurements of the luminosity distance and the estimated
$D_{\scriptstyle A}$ from FRIIb radio galaxies (Daly \& Djorgovski
2003) and ultra compact radio sources (Gurvitz 1999; Lima \&
Alcaniz 2000, 2002; Santos \& Lima 2008) in order to test possible new physics signatures 
based on the following expression:
\begin{equation}
\nonumber
I. \,\,\,\eta(z) = (1+z)^{\beta-1} \exp\left[\gamma \int_0^z
\frac{dz'}{E(z')(1+z')^{\delta}}\right], \label{3param}
\end{equation}
where  $E(z') \equiv H(z')/H_0$ is
the dimensionless Hubble parameter ($H_0$ is the Hubble constant). Note that for arbitrary
values of $\delta$, the strict validity of the DD relation
corresponds to $(\beta, \gamma) \equiv (1,0)$.  By marginalizing on
$\Omega_M$, $\Omega_{\Lambda}$ and Hubble parameters, they found a
$2\sigma$ violation caused by excess brightening of SNIa at $z >
0.5$. It was also argued that such an effect would be associated  to
lensing magnification bias.

De Bernardis, Giusarma \& Melchiorri (2006) also searched for
deviations of the standard CDDR by using the angular diameter distances  from
galaxy clusters provided by the sample of Bonamente {\it et al.}
(2006). By assuming $\eta(z) = constant$, they obtained  a non
violation of CDDR in the framework of the cosmic
concordance $\Lambda$CDM model. Later on, Avgoustidis et al.\ (2009, 2010) working in the context of a flat $\Lambda$CDM model
also adopted an extended CDDR expressed as
\begin{equation}\nonumber
 II. \,\,\,\eta(z) = (1+z)^{\epsilon}.
\end{equation}
The above deformation function is a particular case ($\beta = 1+\epsilon, \gamma=0$)
of the general expression adopted by Basset and Kunz (2004). In their analysis, the recent SNe Ia
data as compiled by Kowalski et al. (2008) were combined with the
latest measurements of the Hubble expansion at redshifts in the
range $0<z<2$ (Stern et al.\ 2010), and the free parameter was constrained to be
$\epsilon=-0.04_{-0.07}^{+0.08}$ (2$\sigma$). More recently, such a parametrization has also been adopted by Khedekar \& Chakaborti (2011) in their studies connecting the Tolman test and CDDR using the redshifted 21 cm wavelength from disk galaxies. It was also argued that future data from the planned  Square Kilometer Array (SKA) my provide the best test to detect any violation of the cosmic distance duality relation. 

In a series of papers,  Holanda, Lima \&  Ribeiro (2010, 2011, 2011a) also explored a different route to test the CDDR based on the following deformation functions:
\begin{equation}\label{param}
III. \,\,\,\eta (z) = 1 + \eta_{0} z, \, \, \, \, \, \,IV.\,\,\, \eta
(z) = 1 + \eta_{0}z/(1+z).
\end{equation}
\noindent   A basic difference between the first and  the second parameterization is that the later includes a
possible epoch dependent correction which avoids the divergence at
extremely high z. At low redshifts, when second order terms are neglected, the second parametrization reduces to the first one, that is, $\eta(z) \simeq 1 + \eta_0 z$. 
Such one-parametric formulas are also interesting
because in the limit of extremaly low redshifts ($z<<1$), one finds $\eta (z) = 1$ as should be expected since $D_{L} = D_{A}$  at this limit (see also parametrization II). 
In addition,  for a given data set, the likelihood of $\eta_0$ must be peaked at
$\eta_0=0$, in order to satisfy the standard duality relation.  
In this context, by taking the SNe Ia from
Constitution data (Hicken et al. 2009) and galaxy clusters samples compiled by De Fillipis et al. (2005) and
Bonamente et al. (2006), a direct test of the CDDR was acomplished (Holanda et al. 2010).  
As an extra bonus, the consistency between the strict validity of CDDR and the assumptions about the geometry based on
elliptical and spherical $\beta$ models was detailedly discussed.  The sphericity assumption for the cluster
geometry resulted in a larger incompatibility with the validity of
the duality relation in comparison with an isothermal non spherical
cluster geometry. More recently, such expressions were adopted by Gon\c{c}alvez et al. (2011) in their studies on the validity of CDDR by including data 
from X-ray gas mass fraction of galaxy clusters.

In this connection, it is also worth mentioning that Li, Wu \& Yu
(2011) rediscussed this independent cosmological test by using the
latest Union2 SNe Ia data and the angular diameter distances from
galaxy clusters thereby obtaining a more serious violation of the standard duality expression.
In a simultaneous but independent work, Nair, Jhingan \& Jain (2011) also investigated the strict validity of
the CDDR by using the latest Union2 SNe Ia data and the angular
diameter distances from galaxy clusters, FRIIb radio galaxies and
mock data. As an attempt to determine a possible
redshift variation of the CDDR relation, they proposed six different (one and two indexes)
parametrizations including, as particular cases, the ones adopted
by Holanda et al.\ (2010, 2011a). As physically expected, their
results depend both on the specific parametrization and the considered data
sample. In particular, they conclude that the one index
parametrizations, namely: $\eta_{V}=\eta_{8}/(1 + z)$ and
$\eta_V=\eta_{9} \exp \{ [z/(1 + z)]/(1 +z) \}$, does not support
the CDDR relation  for the given data set. Meng, Zhang \& Zhan (2011) also reinvestigated the model
independent test by comparing two different methods and several
parametrizations (one and two indexes) for $\eta(z)$.
Their basic conclusion is  that the
triaxial ellipsoidal model is suggested by the model independent
test at 1$\sigma$ while the spherical $\beta$ model can only be accommodated
at $3\sigma$ confidence level thereby agreeing with the results earlier derived  by Holanda et al. (2010).

It should be stressed that all the  above described attempts to test the CDDR  have been carried
out based on a phenomenological approach.  Usually, it is also not clear whether the nonstandard relation is the result of a modified luminosity distance or whether it should be associated to an extended angular diameter distance or both (see, however, Basset and Kunz 2004). At this point, one may also ask whether such expressions to $\eta(z)$ can be derived from a more fundamental approach. In the affirmative case, it is also important to study their consequences for the present accelerating stage of the Universe. 

In this letter we consider both questions. Firstly, we show how any deformed CDDR can be derived by analysing
possible theoretical modifications on the luminosity distance without refraction effects.  Analytical expressions for the dimensionless cosmic absorption parameter describing the above four parametrizations will be explicitly obtained in the framework of Gordon's optical metric as developed by Chen and Kantowski (2009a, 2009b) to include cosmic absorption. It will be also explicitly assumed that the angular diameter distances are not modified because their measurements involve only standard rulers and angular scales, and, more important, possible refractive effects have been neglected.  Apart such hypotheses, the approach discussed here is quite general and can be applied for any deformation function, $\eta(z)$. Secondly, we also apply our results for the latest SNe Ia data. As we shall see, the modified luminosity distance can accomodate the observed supernova dimming even for a non-relativistic cold dark matter (CDM) Einstein - de Sitter model ($\Omega_M=1$). The validity of the $\Lambda$CDM description
is obtained only in the extreme limit of perfect cosmic transparency (negligible cosmic absorption).

\section{Luminosity Distance and Duality Relation}\label{sec:Sample Union2.}

The concept of an optical metric was introduced long ago in a seminal paper by Gordon (1923). He proved the existence of a mapping between any solution of the general relativistic Maxwell's equations for a fluid with refraction index $n(x)$ and the vacuum solutions in a related optical spacetime. In Gordon's treatment, only the refraction phenomenon was considered.  More recently, by describing the Maxwell field as a monochromatic wave, Chen and Kantowski (2009a, 2009b) generalized
such treatment in order to include the possibility of an absorption phenomenom in the Universe.  As it appears, the presence of such an
effect breaks naturally the validity of the standard distance duality relation
as given by Eq. (1)  since the light absorption violates the photon
number conservation law. In this case, they prove that the luminosity distance 
takes the following form:
\begin{equation}\label{eq:dLtau}
D_L(z) =e^{\tau/2} D_L^{S}(z)= {e^{\tau/2} (1+z) \over
H_{0}}\int_0^z {dz'\over E(z')} ,
\end{equation}
where  $\tau$ denotes the optical depth associated to the cosmic
absorption of the Universe and the superscript S specify the
standard luminosity distance (no absorption) for which the Universe is assumed to be transparent. It is worth noticing that effects coming from a possible new physics like the interaction between photons and dark matter as discussed by Basset and Kunz (2004) are assumed to be negligible (no new physics takes place). For a a spatially homogeneous and nondispersive (grey) absorption, the quantity $\tau$  as derived by Chen \& Kantowski (2009a) reads
\begin{equation}\label{tau}
\tau(z) = \int_0^z {\alpha_{\ast}dz'\over {(1+z') E(z')}},
\end{equation}
where $\alpha_{\ast}$ is the dimensionless cosmic absorption parameter
($\alpha/H_{0}$ in the notation of Chen \& Kantowski). In the above expression the refractive index was fixed to unity (n(z)=1), and, therefore, any possible refraction effect has been neglected.  This is an important point, since in this case the standard angular diameter distances are not modified. 
  
Let us now discuss how the several functions $\eta(z)$ defining the  deformed CDDR introduced in an ad-hoc way
can be related with the dimensionless cosmic absorption parameter, $\alpha_{*}(z)$.  As one may check, by inserting
expression (6) into (2),  we obtain that  the cosmic optical depth  and the deformation function
must be  related by the simple expression $e^{\tau/2}=\eta(z)$. In addition, this also means that any smooth $\eta(z)$ deformation function defines an effective cosmic absorption parameter given by:

\begin{equation}
\alpha_{*}(z) = \frac{2 \eta'(z)(1+z)E(z)}{\eta(z)},
\end{equation}
where a prime denotes derivative with respect to the reshift. Such  a relation uniting the dimensionless cosmic absorption parameter and the deformation function, usually introduced by hand in the distance duality relation, is one of the main results of this section. When $\eta(z)$ is constant the cosmic absorption $\alpha_{*}(z)$ is identically null, and, therefore, $\tau = 0$. As remarked earlier, this also means that only the standard relation  as determined by Etherington (1933) is possible at this limit, that is, $\eta(z) = 1$ (note that $\tau$ and $\eta$ are related by an exponential function).  As one may check, the four deformed distance duality relations can analytically be expressed in terms of the dimensionless absorption parameter as:

I. Basset and Kunz (2004):\,\, $\alpha_{*}(z) = 2[(\beta-1)E(z) + \gamma(1+z)^{1-\delta}],$
\vspace{0.2cm}
 
II. Avgoustidis {\it{et al.}} (2009): \, \, $\alpha_{*}(z) = 2\epsilon E(z),$
\vspace{0.2cm}

III. Holanda et al. (2010): \,\, $\alpha_{*}(z) = \displaystyle{\frac{2\eta_{0} (1+z)E(z)}{1 + \eta_0 z}},$
\vspace{0.2cm}

IV. Holanda et al. (2010): \,\, $\alpha_{*}(z) = \displaystyle{\frac{2\eta_{0} E(z)}{1 + z + \eta_0 z}}.$
\vspace{0.2cm}\\
We see that the dimensionless Hubble parameter, $E(z)$, and the free parameters appearing in the deformation functions define completely the cosmic absorption parameter as introduced by Chen and Kantowski (2009a, 2009b). 

At this point some comments are in order. Initially, we notice that in the limiting case $\gamma=0$ and $\beta = \epsilon +1$, the first
expression for the cosmic absorption reduces to the second one. This should be expected since at such a limit the general deformation function of
Basset and Kunz (2004) reduces to the one proposed by Avgoustidis
(2009).  Note also that such expressions are satisfied for all Friedmann-Robertson-Walker (FRW)
geometries  (arbitrary values of $\Omega_k$) and energetic content (baryons, dark
matter, and dark energy). Since we are describing absorption, ($\alpha_* > 0$) this means that the parameters $\epsilon$ and $\eta_0$ must be positive.


\section{Extended Luminosity Distance and Supernova Dimming}

Nowadays, constraints based on SNe Ia data are considered the best
method for studing the cosmic expansion history at $z < 1.5$. Let us now confront the extended luminosity distance including
absorption as an optical metric phenomenon with the latest
Supernova data. In our subsequent analyzes we consider only the flat $\Lambda$CDM model for which the dimensionless Hubble parameter,  $E(z)$, takes the following form: 

\begin{equation}
E(z) = \sqrt{\Omega_{M}(1+z)^3+\Omega_{\Lambda}}, 
\end{equation}
where $\Omega_{\Lambda}= 1 - \Omega_M$.

To begin with, let us consider the Union2 supernova sample which is formed by 557 measurements of distance moduli from Sne Ia 
as compiled by Amanullah {\it{et al.}} (2010). In order to avoid effects from Hubble bubble, only 506
supernovae with redshifts greater than $cz = 7000 km/s$ were selected (Conley {\it{et al.}} 2007, Kessler {\it{et al.}}
2009, Sinclair {\it{et al.}} 2010). As widely known, the SNIa Union2 data are
obtained by adding new datapoints (including the high redshift
SNIa) to the original Union data (Kowalski et al. 2008). For this enlarged sample, a number of refinements to the
original Union analysis chain has been done, in particular, the relative importance of systematic effects was higlighted (in this connection see also Sullivan et al. (2011) to the Legacy Survey sample (SNLS3)). 

In our statistical analysis we consider a maximum likelihood determined by a
$\chi^{2}$ statistics 
\begin{equation}
\chi^2(\mathbf{p}) = \sum_{SNIa} {[\mu_{obs}(z_i; \mathbf{p})-
\mu_{th}{(z_i)}]^2 \over \sigma_{obs,i}^2}, \label{chi2}
\end{equation}
where $\mu_{th}(z_i)=5\log_{10}D_L(z_i, \mathbf{p}) +\mu_0$ is the
theoretical distance moduli, $\mu_0=25-\log_{10} H_0$, $D_L$ is the
luminosity distance from Eq. (6), $\sigma_{obs,i}$  is the
uncertainty in the individual distances (including systematic errors), and the complete set of
parameters is given by $\mathbf{p} \equiv (H_{0},
\Omega_{M},\alpha_{*})$.  It should be stressed that for the considered SNe Ia subsample we have combined the  statistical plus
systematic errors in quadrature as compiled in Table 7 of Amanullah {\it et al.} (2010) and neglected $\sigma^{2}_{lc}$. We have also marginalized on the Hubble constant.

 In Fig. 1(a) we show the contours on the $\Omega_M-\alpha_{*}$ plane corresponding to a flat
$\Lambda$CDM model and by considering that the absorption 
$\alpha_{*}$ is constant. As indicated, the shadow lines are cuts in the regions of 68.3\%, 95.4\% and 99.7\% of probability. The constraints on the free parameters are restricted to $0.0 \leq \alpha_{*} \leq 1.55$ and $0.21 \leq \Omega_M \leq 1.0$ at $2\sigma$ of statistical confidence  while to the latter we have obtained  $0.0 \leq \alpha_{*} \leq 1.46$ and $0.22 \leq \Omega_M \leq 1.0$ at $2\sigma$ of statistical confidence.  In the absence of absorption  ($\alpha_{*}=0$) the  limit on the density parameter is $0.21 \leq \Omega_M \leq 0.34$ (concordance model) while for $\Omega_M=1$ (Einstein de Sitter), we find that the absorption parameter lies on the interval $1.20 \leq \alpha_{*} \leq
1.54$ ($2\sigma$). Note that the positiveness of $\alpha_{*}$ (absorption), implies that a pure de
Sitter model ($\Omega_{\Lambda}=1, \Omega_M=0$) is not allowed by such data. The best fit
to the free parameters are $\Omega_M=0.27, \alpha_{*}=0.0$ with
$\chi^2_{min}=330.5$.

In Fig. 1(b) we display the likelihood
distribution functions, $e^{-\chi^{2}/2}$  of $\alpha_{*}$,
for the Einstein-de Sitter model ($\Omega_M=1$) with
constant absorption. The  blue 
curve includes statistical + systematic errors, but, for comparison, we have also shown the red curve with only statistical errors. The  upper and lower horizontal  lines correspond to $68.3\%$, and $95.4\%$ c.l., respectively. 
By marginalizing on the Hubble parameter we obtain $\alpha_{*}=1.38 \pm 0.08 (0.15)$ with 1$\sigma$ (2$\sigma$) of probability and
$\chi^2_{min}=331.3$. In this model the Universe is always decelerating since $q(z)= q_0= 1/2$. Instead of dark 
energy we have a cosmic medium whose absorbing properties is quantified by the dimensionless parameter $\alpha_{*} \simeq 1.4$ 
which is responsible for the SNe Ia dimming. The dimensional absorption parameter, $\alpha = \alpha_*H_0 \sim 10^{-4} Mpc^{-1}$, is 
nearly the same one previously obtained  by  Chen and Kantowski (2009a).

Let us now consider the deformation function, $\eta(z) = (1+z)^{\epsilon}$, as adopted  by Avgoustidis {\it{et al.}} (2009). As shown in the previous section, the associated cosmic absorption parameter in this case is $\alpha_{*}(z) = 2\epsilon E(z)$. In order to simplify the notation, in what follows we consider the parameter, $\alpha_0 = 2\epsilon$. 

In Figure 2(a) we display the corresponding plots for the ($\Omega_M,
\alpha_{0}$) plane to the case of variable absorption, $\alpha_{*}= \alpha_0 E(z)$. As in Fig. 1(a) the analysis includes statistical plus systematic errors.  
The confidence region (2$\sigma$) in this plane is defined by $0.0 \leq
\alpha_{0} \leq 0.92$ and $0.20 \leq \Omega_M \leq 1.0$. Again, the best fit is the $\Lambda$CDM  with 
best fit $\Omega_M=0.27, \alpha_0= 0.0$ with
$\chi^2_{min}=330.5$.

 In Fig. 2(b)  we display the likelihood distribution functions for
$\alpha_{0}$. Blue and red curves  correspond to the same analysis of Fig. 1(b), that is, with and without systematics. To the latter case,  we obtain that $\alpha_{0}
=0.88 \pm 0.05 (0.10)$ with 1$\sigma$ (2$\sigma$) of confidence
level. Once again we have $q(z)= q_0= 1/2$ in the presence of the
absorption, but with $\chi^2_{min}=335.9$. 

It should be stressed  that the above results for the $\Omega_M - \alpha_{*}$ and $\Omega_M - \alpha_{0}$ planes,  based on
a subsample of the Union2 SNe  Ia data, are strongly suggesting that one of the following possibilities must be true: 

(i) The Universe has an extra dark energy component ($\Lambda$) which provides the fuel for the observed accelerating stage
$(q_0<0, \alpha_{*}=0)$.

(ii) We live in a decelerating Einstein-de Sitter Universe ($\Omega_M=1$) endowed with a cosmic absorption mechanism which is responsible for the dimming of the distant Supernovae Ia  ($q_0 = 1/2, \alpha_{*} \neq 0$). 

(iii) We are not living in the extreme cases above mentioned, that is, both  phenomena (cosmic oppacity and $\Lambda$) are partially operating in the observed Universe.

In conclusion, by adopting the
Gordon-Chen-Kantoswski description we have shown how to obtain analytical expressions for any deformed distance duality relation (parametrized by  $\eta(z)$ function) in terms of the cosmic absorption parameter and vice-versa.  Four especific examples of $\eta(z)$ functions, recently proposed in the literature, were explicitly considered. The associated absorption parameter for each one was explicitly determined and some of its  physical implications also were discussed in detail. Two different scenarios with cosmic absorption were proposed and their  free parameters constrained  by using the SNe Ia data (Amanullah et al. 2010). Working in the in the framework of a flat $\Lambda$CDM model we have found that  both scenarios ($\alpha_{*}= \alpha_{0}E(z)$  and $\alpha_{*}$ = constant)  are also compatible with a decelerating Einstein-de Sitter Universe (see Figures 1a and 2a).  Finally, as remarked before,
in the present study we have neglected  any effect due to a possible light refraction phenomenon as described in the original Gordon's description. The general case it will be discussed in a forthcoming communication.

\begin{acknowledgments}
\noindent JASL and VTZ are partially supported by CNPq and FAPESP (Brazilian
Research Agencies). JVC is supported by CAPES.
\end{acknowledgments}

\clearpage


\begin{figure}[h!]
    \centering
        \includegraphics[width=0.45\linewidth]{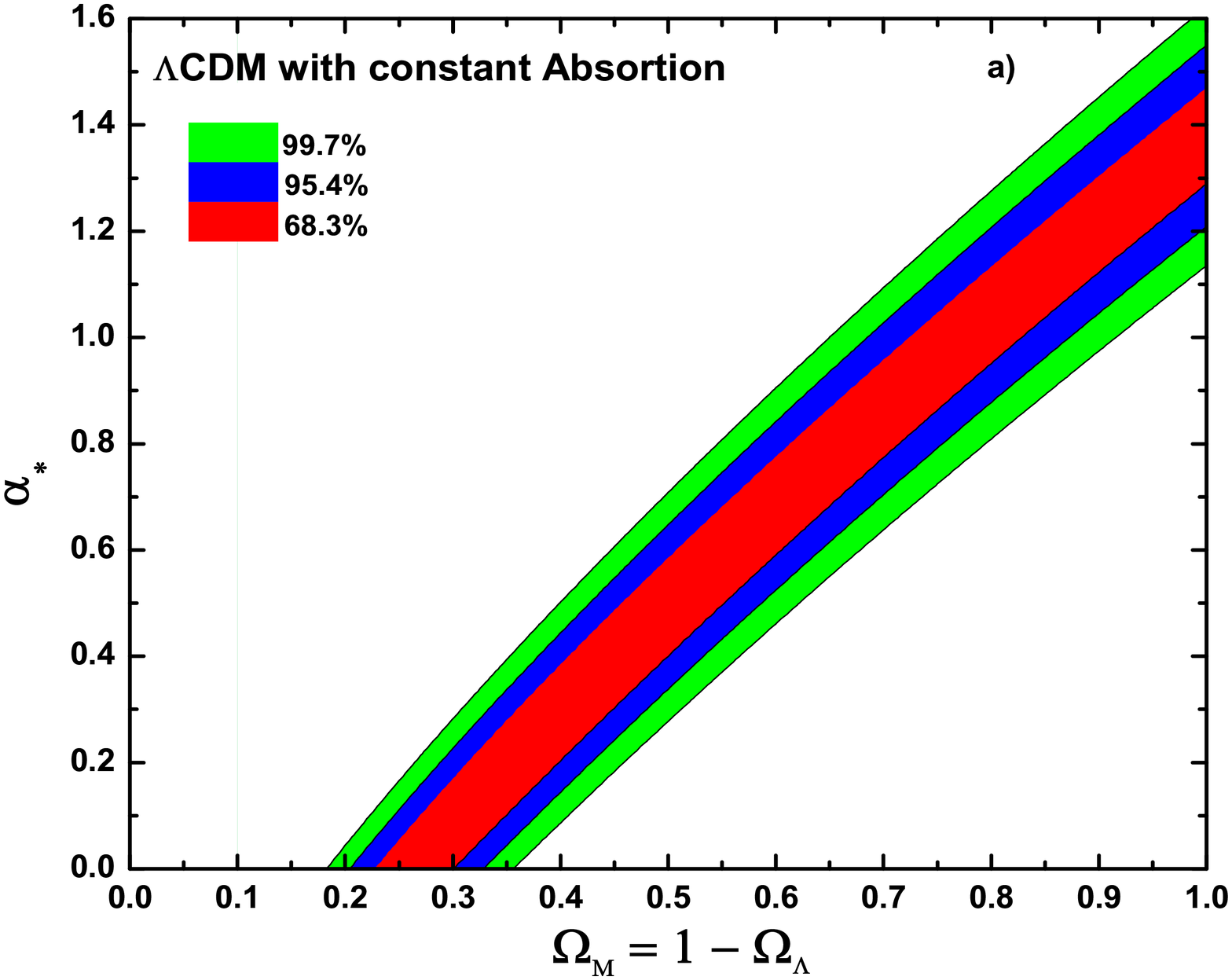}
        \includegraphics[width=0.45\linewidth]{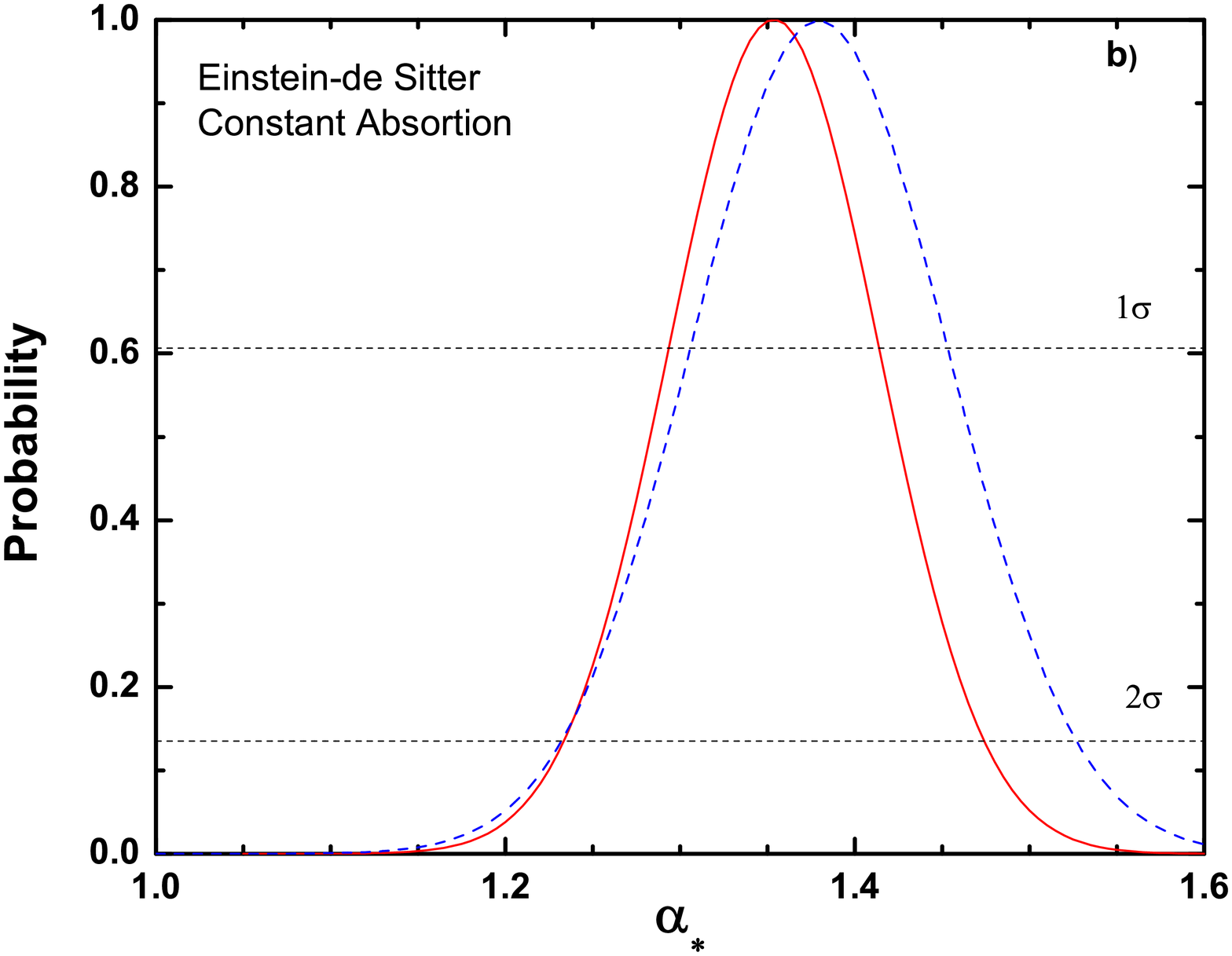}
    \caption{{\bf a)} Contours in the
$\Omega_M - \alpha_{*}$ plane for 506 supernovae data by considering a flat $\Lambda$CDM model with constant absorption
parameter, $\alpha_{*}$. Note that the Einstein-de Sitter model is now allowed by such data. {\bf b)}  The likelihood function of the constant absorption $\alpha_{*}$  for the Einstein-de Sitter model ($\Omega_M = 1$).  The red line takes into account only statistical while the blue curve includes both errors (statistical + systematic). The  upper and lower horizontal lines correspond to $68.3\%$ ($1\sigma$), and $95.4\%$ ($2\sigma$) c.l., respectively. The free parameter is constrained to $\alpha_{*} = 1.38 \pm 0.08$ (statistical + systematic, $1\sigma$ c.l.).}
\end{figure}

\begin{figure}[h!]
    \centering
        \includegraphics[width=0.45\linewidth]{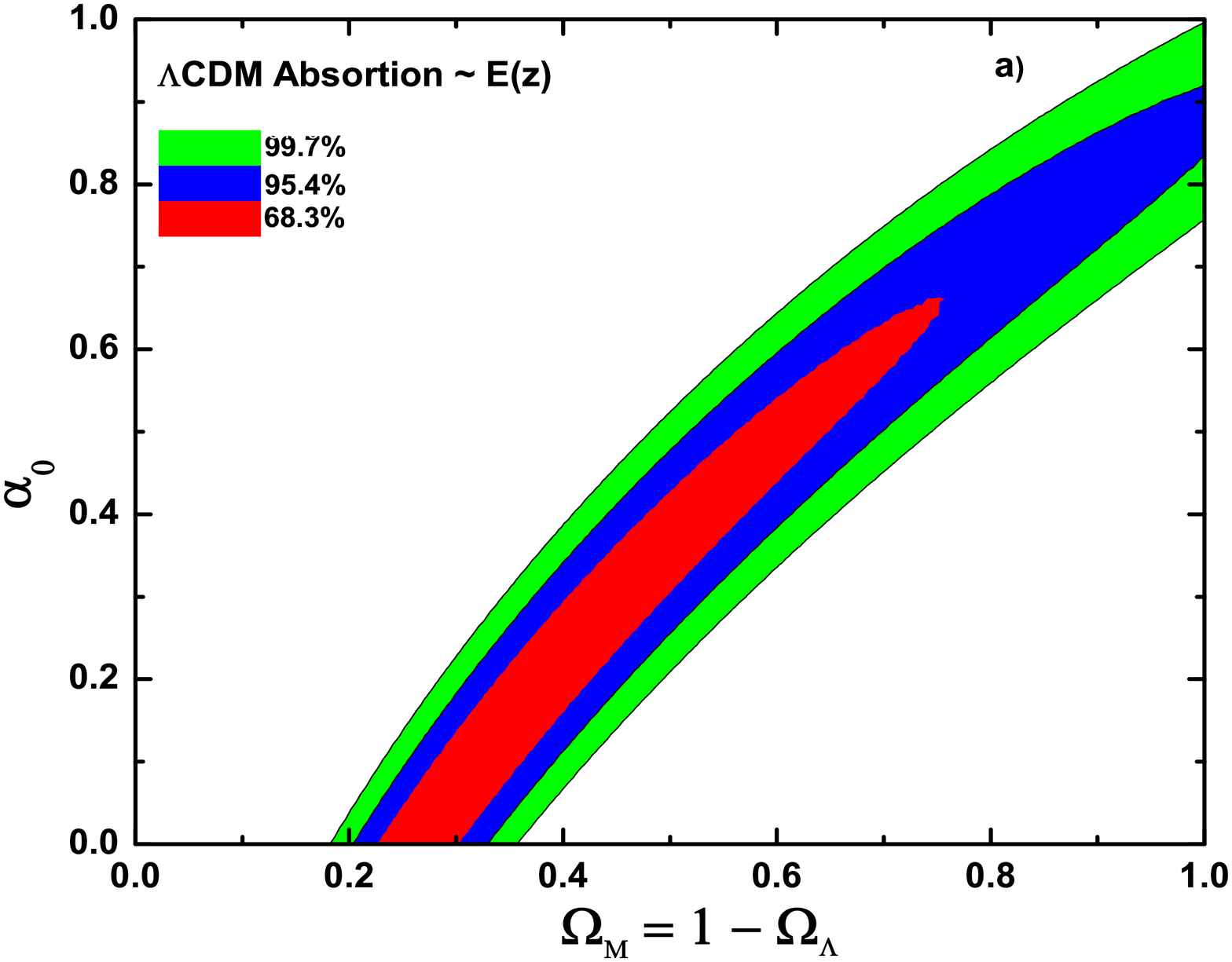}
        \includegraphics[width=0.45\linewidth]{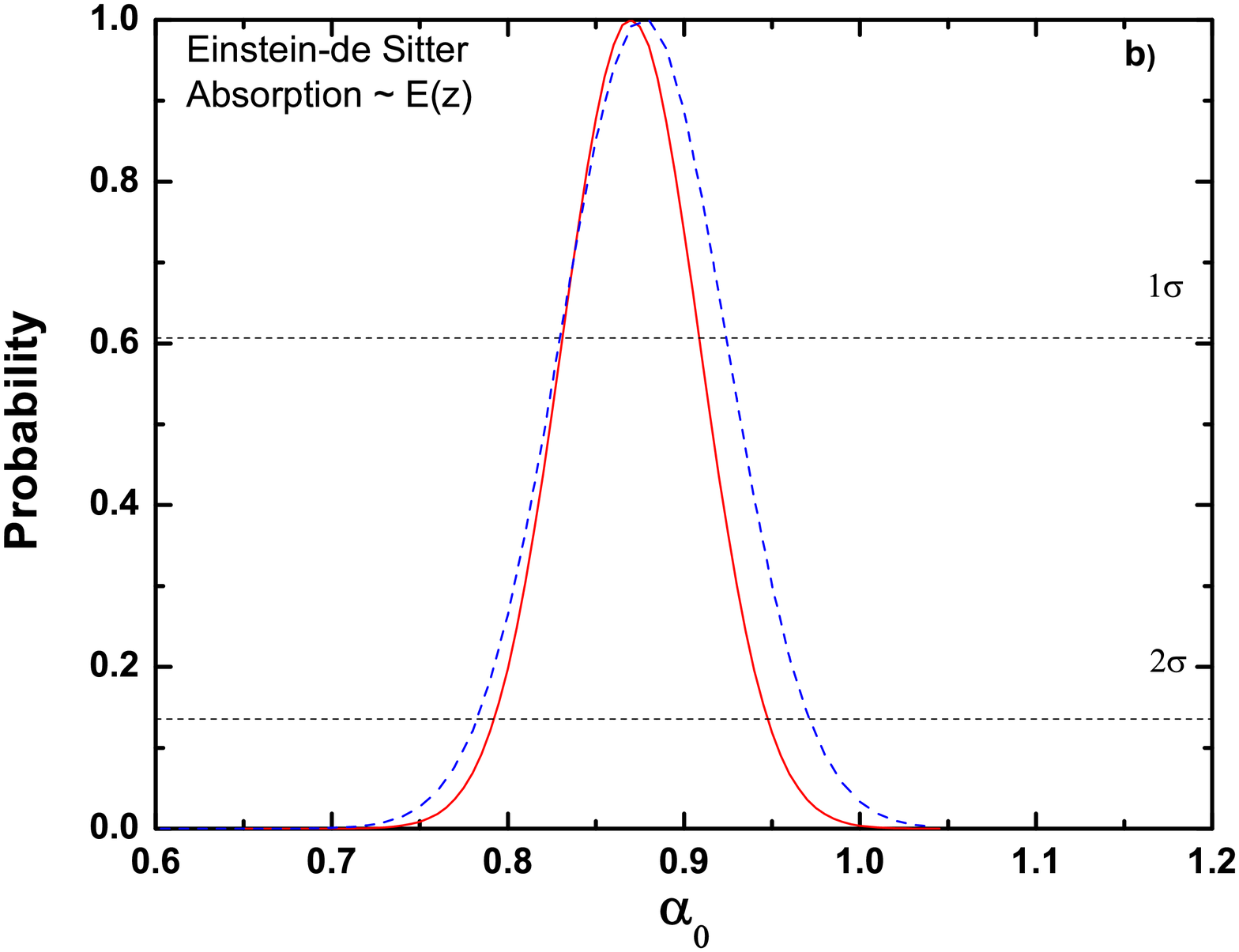}
    \caption{{\bf a)} Contours in the
$\Omega_M - \alpha_{0}$ plane for the same 506 supernovae data by considering a flat $\Lambda$CDM model with
an epoch-dependent absorption, $\alpha_{*}=\alpha_0 E(z)$, as fixed by the Avgoustidis et al. (2009) 
deformation function (see sections I and II). The best fit to the pair of free parameters is given by ($\Omega_M, \alpha_{0}$) = ($0.27,0.0)$. {\bf b)} The likelihood functions of $\alpha_{0}$ in the case of flat Einstein-de Sitter model.  The free parameter is constrained to $\alpha_{0} = 0.88 \pm 0.05$ (statistical + systematic, $1\sigma$ c.l.). }
\end{figure}

\end{document}